\documentclass{article}
\usepackage{graphicx}
\usepackage{epsfig}
\begin{document}
\baselineskip=16pt
\begin{center}
{\Large\sl An Alternative Model for the Tidal Evolution of the Earth-Moon-Sun System}\\
\vspace{2cm} \textbf{Arbab I. Arbab} \footnote{E-mail:
aiarbab@uofk.edu} \\
\vspace{1cm} {Department of Physics, Faculty of Science,
University of Khartoum, P.O. Box 321, Khartoum 11115, Sudan}
\end{center}
\vspace{2cm} {\small We have found that the expansion of the
universe has immense consequences on our local systems. We present
a model based on cosmic expansion that fits well with observation.
The close approach problem inflicting tidal theory is averted in
this model. We have shown that the astronomical and geological
changes of our local
systems are of the order of Hubble constant.}\\
\\
Key words: Past rotation, Tidal theory, Earth-Moon system,
Acceleration, Geologic time,
 angular momentum, secular changes
 \vspace{.5in}
\section{Introduction}
The study of the  Earth Moon-Sun system is very important and
interesting. Newton's laws of motion can be applied to such a
system and good results are obtained. However, the correct theory
to describe the gravitational interactions is the general theory
of relativity. The theory is prominent in describing a compact
system, such as neutron stars, black hole, binary pulsars, etc.
 Einstein theory is applied to study the evolution of the
universe. We came up with some great discoveries related to the
evolution of the universe. Notice that the Earth-Moon system is a
relatively old system (4.5 billion years) and would have been
affected by this volution.Firstly, the model predicts the right
abundance of Helium in the universe during the first few minutes
after the big bang. Secondly, the model predicts that the universe
is expanding and that it is permeated with some relics photons
signifying a big bang nature. Despite this great triumphs, the
model is infected with some troubles. It is found the age of the
universe determined according to this model is shorter than the
one obtained from direct observations. To resolve some of these
shortcomings we propose a model in which vacuum decays with time
couples to matter. This would require the gravitational and
cosmological constant to vary with time too.  To our concern we
have found that the gravitational interactions in the Newtonian
picture can be applied to the whole universe provided we make the
necessary arrangement. First of all, we know beforehand that the
temporal behavior is not manifested in the Newton law of
gravitation. It is considered that gravity is static. We have
found that instead of considering perturbation to the Earth-Moon
system, we suggest that these effect can be modelled with having
an effective coupling constant ($G$) in the ordinary Newton's law
of gravitation. This effective coupling takes care of the
perturbations that arise from the effect of other gravitational
objects. At the same time the whole universe is influenced by this
setting. We employ a cosmological models that describe the present
universe and solves for many of the cosmological problems. To our
surprise the present cosmic acceleration can be understood as a
counteract due to an increasing gravitational strength. The way
how expansion of the universe affects our Earth-Moon system shows
up in changing the length of day, month, distance, etc. These
changes are found in some biological and geological systems. In
the astronomical and geological frames changes are considered in
terms of tidal effects induced  by the Moon on the Earth. However,
tidal theory runs in some serious difficulties when the distance
between Earth and Moon is extrapolated backwards. The Moon must
have been too close to the Earth a situation that has not been
believed to have happened in our past. This will bring the Moon
into a region that will make the Moon rather unstable and the
Earth experiencing a big tide that would have melt the whole
Earth. We have found that one can account for this by an
alternative consideration in which expansion of the universes is
the main cause.
\section{Tidal Theory}
We know that the Earth-Moon system is governed by Kepler's laws.
The rotation of the Earth in the gravity field of the Moon and Sun
imposes periodicities in the gravitational potential at any point
on the surface. The most obvious effect is the ocean tide which is
greater than the solid Earth tide. The potential arising from the
combination of the Moon's gravity and rotation with orbital
angular velocity ($\omega_L$) about the axis through the common
center of mass is (Stacey, 1977)
\begin{equation}\label{}
    V=-\frac{Gm}{R'}-\frac{1}{2}\omega^2_Lr^2
\end{equation}
where $m$ is the mass of the Moon, and from the figure below one
has
\begin{equation}\label{}
    R'^2=R^2+a^2-2a R\cos\psi\ ,\qquad
    r^2=b^2+a^2\sin^2\theta-2ab\cos\psi\ ,
\end{equation}
where $\cos\psi=\sin\theta\cos\lambda$, $b=\frac{m}{ M+m}R$, $a$
is the Earth's radius.
\begin{center}
\includegraphics{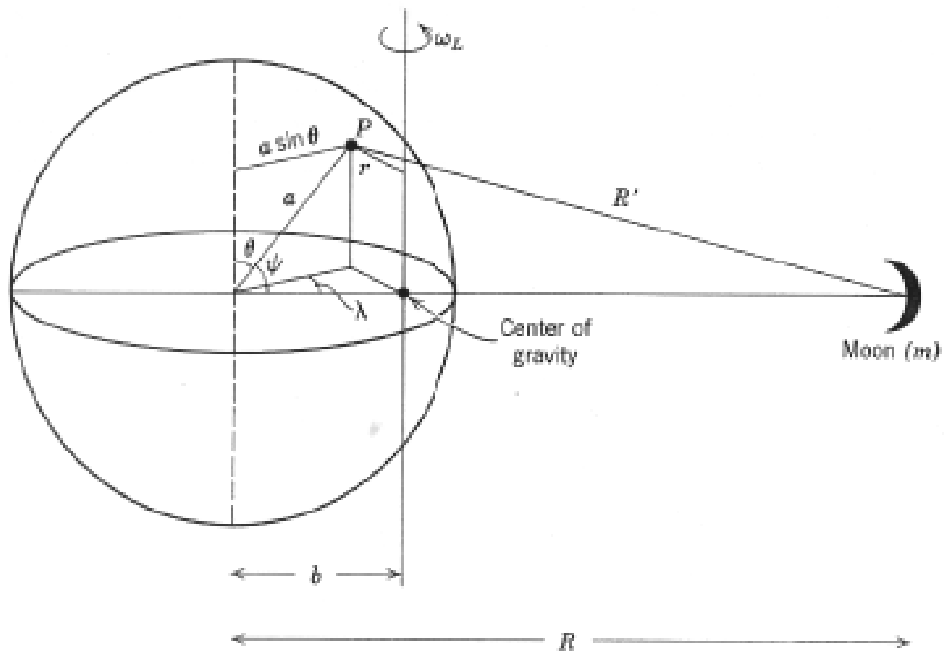}
\end{center}
\centerline{\emph{The geometry of the calculation of the tidal
potential of the Moon and a point $P$ on the Earth's surface.}}
\vspace{1cm} From Kepler's third law one finds
\begin{equation}\label{}
    \omega^2_LR^3=G(M+m)
\end{equation}
where $M$ is the Earth's mass, so that one gets for $a<<R$
\begin{equation}\label{}
    V=-\frac{Gm}{R}\left(1+\frac{1}{2}\frac{m}{M+m}\right)-
\frac{Gma^2}{R^3}(\frac{3}{2}\cos\psi-\frac{1}{2})-\frac{1}{2}\omega^2_La^2\sin^2\theta
\end{equation}
The first term is a constant that is due to the gravitational
potential due to the Moon at the center of the Earth, with small
correction arising from the mutual rotation. The second term is
the second order zonal harmonics and represents a deformation of
the equipotential surface to a prolate ellipsoid aligned with the
Earth Moon axis. Rotation of the Earth is responsible for the
tides. We call the latter term tidal potential and define it as
\begin{equation}\label{}
    V_2=-\frac{Gma^2}{R^3}(\frac{3}{2}\cos\psi-\frac{1}{2}).
\end{equation}
The third term is the rotational potential of the point $P$ about
an axis through the center of the Earth normal to the orbital
plane. This does not have a tidal effect because it is associated
with axial rotation and merely becomes part of the equatorial
bulge of rotation. Due to the deformation an additional potential
$k_2V_2$ ($k_2$is the Love number) results, so that at the
distance ($R$) of the Moon the form of the potential due to the
tidal deformation of the Earth is
\begin{equation}\label{}
    V_T=k_2V_2=k_2\left(\frac{a}{R}\right)^3=-\frac{G m a^5}{R^6}(\frac{3}{2}\cos\psi-\frac{1}{2}).
\end{equation}
We can now identify $\psi$ with $\phi_2$: the angle between the
Earth Moon line and the axis of the tidal bulge, to obtain the
tidal torque ($\tau$) on the Moon:

\begin{equation}\label{}
    \tau=m\left(\frac{\partial V_T}{\partial \psi}\right)_{\psi=\phi_2}
    =\frac{3}{2}\left(\frac{Gm^2a^5k_2}{R^6}\right)\sin2\phi_2
\end{equation}
The torque causes an orbital acceleration of the Earth and Moon
about their common center of mass; an equal and opposite torque
exerted by the Moon on the tidal bulge slows the Earth's rotation.
This torque must be equated with the rate of change of the orbital
angular momentum ($L$), which is (for circular orbit)
\begin{equation}\label{}
    L=\left(\frac{M}{M+m}\right)R^2\omega_L\ ,
\end{equation}
upon using eq.(3) one gets
\begin{equation}
L=\frac{Mm}{M+m}(GR)^{\frac{1}{2}}\ , \qquad L=
\frac{MmG^{\frac{2}{3}}}{(M+m)^{\frac{1}{3}}}\omega^{-\frac{1}{3}}_L\
.
\end{equation}
The conservation of the total angular momentum of the Earth-Moon
system ($J$) is a very integral part in this study. This can be
described as a contribution of  two terms: the first one due to
 Earth axial rotation ($S=C\omega$) and the second term due to the Moon orbital rotation
($L$). Hence, one writes
\begin{equation}\label{}
J=S+L=C\omega+\left(\frac{Mm}{M+m}\right)R^2\omega_L
\end{equation}
We remark here to the fact that of all planets in the solar
system, except the Earth, the orbital angular momentum of the
satellite is a small fraction of the rotational angular momentum
of the planet. Differentiating the above equation with respect to
time $t$ one gets
\begin{equation}\label{}
\tau=\frac{dL}{dt}=\frac{L}{2R}\frac{dR}{dt}=-\frac{dS}{dt}
\end{equation}

The corresponding retardation of the axial rotation of the Earth,
assuming conservation of the total angular momentum of the
Earth-Moon system, is
\begin{equation}\label{}
    \frac{d\omega}{dt}=-\frac{\tau}{C}
\end{equation}
assuming $C$ to be constant, where $C$ is the axial moment of
inertia of the Earth and its present value is $(C_0=8.043\times
10^{37}\rm kg\ m^{-3}$). It is of great interest to calculate the
rotational energy dissipation in the Earth-Moon system. The total
energy ($E$) of the Earth - Moon system is the sum of three terms:
the first one due to axial rotation of the Earth, the second is
due to rotation of the Earth and Moon about their center of mass,
and the third one is due to the mutual potential energy.
Accordingly, one has
\begin{equation}\label{}
    E=\frac{1}{2}C\omega^2+\frac{1}{2}R^2\omega^2_L\left(\frac{Mm}{M+m}\right)-\frac{GMm}{R}
\end{equation}
and upon using eq.(3) become
\begin{equation}\label{}
    E=\frac{1}{2}C\omega^2-\frac{1}{2}\frac{GMm}{R}\ ,
\end{equation}
Thus
\begin{equation}\label{}
    \frac{dE}{dt}=C\omega\frac{d\omega}{dt}-\frac{1}{2}\frac{GMm}{R^2}\frac{dR}{dt}
\end{equation}
using eqs. (8), (11) and (12) one gets
\begin{equation}\label{}
\frac{dE}{dt}=-\tau(\omega-\omega_L)
\end{equation}

\section{Our Alternative Model}
Instead of using the tidal theory described above, we rather use
the ordinary Kepler's and Newton law of gravitational. We have
found that the gravitation constant $G$ can be written as
\begin{equation}\label{}
    G_{\rm eff.}=G_0 f(t)
\end{equation}
where $f(t)$ is some time dependent function that takes care of
the expansion of the universe. At the present time we have
$f(t_0)=1$. It seems as if Newton's constant changes with time. In
fact we have effects that act as if gravity changes with time.
These effects could arise from any possible source (internal or
external to Earth). This variation is a modelled effect due to
perturbations received from distant matter. This reflects the idea
of Mach who argued that distant matter affects inertia. We note
here the exact function $f(t)$ is not known exactly, but we have
its functional form. It is of the form $f(t)\propto t^n$, where
$n>0$ is an undetermined constant which has to be obtained from
experiment (observations related to the Earth-Moon system). Unlike
Dirac hypothesis in which $G$ is a decreasing function of time,
our model here suggests that $G$ increases with time. With this
prescription in hand, the forms of Kepler's and Newton's laws
preserve their form and one does not require any additional
potential (like those appearing in eqs.(5) and (6)) to be
considered. The total effect of such a potential is incorporated
in $G_{\rm eff.}$. We have found recently that (Arbab, 1997)
\begin{equation}\label{}
    f(t)=\left(\frac{t}{t_0}\right)^{1.3}
\end{equation}
where $t_0$ is the present age of the universe, in order to
satisfy Wells and Runcorn data (Arbab, 2004).
\subsection{The Earth-Sun System} The orital angular momentum of
the Moon is given by
\begin{equation}\label{}
 L_S=\left(\frac{M}{M+M_\odot}\right)R^2\omega_L
\end{equation}
or,
\begin{equation}\label{}
L_S=\left(\frac{M M_\odot}{M+M_\odot}\right)(G_{\rm
eff}R)^{\frac{1}{2}}\ ,\qquad L_S
=\left(\frac{MM_\odot}{M+M_\odot}\right)^{\frac{1}{3}}\left(\frac{G_{\rm
eff}^2}{\Omega}\right)^{\frac{1}{3}}
\end{equation}
where we have replace $G$ by $G_{\rm eff.}$, and $\Omega$ is the
orbital angular velocity of the Earth about the Sun. The length of
the year ($Y$) is given by Kepler's third law as
\begin{equation}\label{}
Y^2=\left(\frac{4\pi^2}{G_{\rm eff}(M_\odot+M)}\right)R_E^3
\end{equation}
where $Y$ is length of the year and $R_E$ is the Earth-Sun
distance. We normally measure the year not in a fixed time but in
terms of number of days. If the length of the day changes, the
number of days in a year also changes. This induces an apparent
change in the length of year. From eqs.(20) and (21) one obtains
the relation
\begin{equation}\label{}
    L_S^3=N_1G_{\rm eff}Y^2
\end{equation}
and
\begin{equation}\label{}
    L^2_S=N_2G_{\rm eff}R_E
\end{equation}
where $N_1, N_2$ are some constants involving ($m, M, M_\odot$).
Since the angular momentum of the Earth-Sum remains constant, one
gets the relation
\begin{equation}\label{}
    Y^2=Y_0\left(\frac{G_{\rm eff}}{G_0}\right)^2
\end{equation}
where $Y$ is measured in terms of days, $Y_0=365.24$ days. Eq.(23)
gives
\begin{equation}\label{}
    R_E=R^0_E\left(\frac{G_0}{G_{\rm eff}}\right)
\end{equation}
where $R^0_E=1.496\times 10^{11}\rm m$. To preserve the length of
year (in terms of seconds) we must have the relation
\begin{equation}\label{}
    D=D_0\left(\frac{G_0}{G_{\rm eff}}\right)^2
\end{equation}
so that
\begin{equation}\label{}
Y_0D_0=YD=3.155\times 10^{7}\rm s
\end{equation}
This fact is supported by data obtained from paleontology. We know
further that the length of the day is related to $\omega$ by the
relation $D=\frac{2\pi}{\omega}$. This gives a relation of the
angular velocity of the Earth about its self of the form
\begin{equation}\label{}
    \omega=\omega_0\left(\frac{G_{\rm eff}}{G_0}\right)^2
\end{equation}
\subsection{The Earth-Moon system} The orbital angular momentum of
the Moon is given by
\begin{equation}\label{}
 L=\left(\frac{M}{M+m}\right)R^2\omega_L
\end{equation}
or,
\begin{equation}\label{}
L=\left(\frac{Mm}{M+m}\right)(G_{\rm eff}R)^{\frac{1}{2}}\ ,\qquad
L =\left(\frac{Mm}{M+m}\right)^{\frac{1}{3}}\left(\frac{G_{\rm
eff}^2}{\omega_L}\right)^{\frac{1}{3}}
\end{equation}
where we have replace $G$ by $G_{\rm eff.}$, and $\omega_L$ is the
orbital angular velocity of the Moon about the Earth. However, the
length of month is not invariant as the angular momentum of the
Moon has not been constant over time. It has been found found by
Runcorn that the angular momentum of the Moon 370 million years
ago (Devonian) in comparison to the present one ($L_0$) to be

\begin{equation}\label{1}
    \frac{L_0}{\L}=1.016\pm 0.003
\end{equation}
The ration of the present angular momentum of the Moon  ($L$) to
that of the Earth ($S$) is given by
\begin{equation}\label{}
    \frac{L_0}{S_0}=4.83
\end{equation}
so that the total angular momentum of the Earth Moon system is
\begin{equation}\label{}
    J=L+S=L_0+S_0=3.4738\times 10^{34}\rm Js
\end{equation}
 Hence, eqs. (28), (29) and (30) become, using eqs.(17) and (18)

\begin{equation}\label{}
   L=L_0\left(\frac{t}{t_0}\right)^{0.44}\
    , \qquad \omega=\omega_0\left(\frac{t_0}{t}\right)^{2.6}\ , \qquad \omega_L=\omega_{0L}\left(\frac{t}{t_0}\right)^{1.3}\
\end{equation}
The length of the sidereal month is given by
\begin{equation}\label{}
    T=\frac{2\pi}{\omega_L}=T_0\left(\frac{t_0}{t}\right)^{1.3}
\end{equation}
where $T_0=27.32$ days, and the synodic month is given by the
relation
\begin{equation}\label{}
    T_{sy}=\left(\frac{T}{1-\frac{T}{Y}}\right)
\end{equation}
 From eqs.(34) one finds
\begin{equation}\label{}
    \omega\ \omega_L^2=\omega_0\ \omega^2_{0L}
\end{equation}
If the Earth and Moon were once in resonance then
$\omega=\omega_L\equiv\omega_c$. This would mean that
\begin{equation}\label{}
    \omega_c^3\ =\omega_0\ \omega^2_{0L}=516.6\times
    10^{-18}\ \rm(rad/s)^3\ , \qquad \omega_c=8.023\times 10^{-6}\rm\ rad/s
\end{equation}
We notice that the present Earth declaration is $5.46\times
10^{-22}\rm rad/s^2$, or equivalently a lengthening of the day at
a rate of 2 milliseconds per century. The increase in Moon mean
motion is $9.968\times 10^{-24}\rm rad/s^2$. Hence, we found that
$\dot \omega=-54.8\ \dot n$. The month increases by 27.88
millisecond/cy. This would mean that both the length of day and
month were equal. They were both equal to a value of about 9
present days. Such a period has not been possible since when the
Earth was formed the month was about 14 present days and the day
was 6 hours! Therefore, the Earth and Moon had never been in
resonance in the past.

Using the eqs.(11) and (34) the torque on the Earth by the Moon is
\begin{equation}\label{}
    \tau=-\frac{dL}{dt}=-\frac{dS}{dt}\ , \qquad \tau=-\tau_0\left(\frac{t}{t_0}\right)^{0.56}\
    ,
\end{equation}
where $\tau_0=3.65\times10^{15}\rm N\ m$. The energy dissipation
in the Earth is given by
\begin{equation}\label{}
    P=\frac{dE}{dt}\ , \qquad \frac{dE}{dt}=\frac{d}{dt}\left(\frac{1}{2}C\omega^2-\frac{1}{2}\frac{G_{\rm eff.}Mm}{R}\right)\ ,
\end{equation}
where $R$, $\omega$ is given by eqs.(30) and (34).

We remark that  the change in  the Earth-Moon-Sun parameters is
directly related to Hubble constant ($H$). This is evident since
in our model (see Arbab, 1997) the Hubble constant varies as
$H=1.11 t^{-1}$. Hence, one may attribute these changes to cosmic
expansion.

Fossils of coral reefs studied by John Wells (1963) revealed that
the number of days in the past geologic time was bigger than now.
This entails that the length of day was shorter in the past than
now. The rotation of the Earth is gradually slowing down at about
2 milliseconds a century. Another method of dating that is popular
with some scientists is tree-ring dating. When a tree is cut, you
can study a cross-section of the trunk and determine its age. Each
year of growth produces a single ring. Moreover, the width of the
ring is related to environmental conditions at the time the ring
was formed. It is therefore possible to know the length of day in
the past from paleontological studies of annual and daily growth
rings in corals, bivalves, and stromatolites. The creation of the
Moon was another factor that would later help the planet to become
more habitable.  When the day was shorter the Earth's spins
faster. Hence, the Moon tidal force reduced the Earth's rotational
winds. Thus, the Moon stabilizes the Earth rotation and the Earth
became habitable. It is thus plausible to say that the Earth must
have recovered very rapidly after the trauma of the Moon's
formation. It was found that circadian rhythm in higher animals
does not adjust to a period of less than 17-19 hours per day. Our
models can give clues to the time these animals first appeared
(945-1366 million years ago).

This shortening is attributed to tidal forces raised by the Moon
on Earth. This results in slowing down the Earth rotation while
increasing the orbital motion of the Moon. According to the tidal
theory explained above we see that the tidal frictional torque
$\tau\propto R^{-6}$ and the amplitude of tides is $\propto
R^{-3}$. Hence, both terms have been very big in the past when $R$
was very small. However, even if we assume the rate
$\frac{dR}{dt}$ to have been constant as its value now, some
billion years ago the Earth-Moon distance $R$ would be very short.
This close approach would have been catastrophic to both the Earth
and the Moon. The tidal force would have been enough to melt the
Earth's crust. However, there appears to be no evidence for such
phenomena according to the geologic findings. This fact places the
tidal theory, as it stands, in great jeopardy. This is the most
embarrassing situation facing the tidal theory.

\section{Velocity-dependent Inertia Model}
A velocity - dependent inertial induction model  is recently
proposed by Ghosh (1985) in an attempt to surmount this
difficulty. It asserts that a spinning body slows down in the
vicinity of a massive object. He suggested that part of the
secular retardation of the Earth's spin and of the Moon's orbital
motion can be due to inertial induction by the Sun. If the Sun's
influence can make a braking torque on the spinning Earth, a
similar effect should be present in the case of other spinning
celestial objects. This theory predicts that the angular momentum
of the Earth ($L'$), the torque ($\tau')$, and distance ($R'$)
vary as
\begin{equation}\label{}
L'=\frac{mM}{(M+m)^{\frac{1}{3}}}G^{\frac{2}{3}}\omega_L^{-\frac{1}{3}}\
, \qquad \tau'=-\frac{L'}{3\omega_L}\dot\omega_L\ , \qquad \dot
R=-\frac{2}{3}\frac{R}{\omega_L}\dot\omega_L
\end{equation}
The  present rate of the secular retardation of the Moon angular
speed is found to be
$\frac{d\omega_L}{dt}\equiv\dot\omega_L\approx 0.27\times
10^{-23}\rm rad \ s^{-2}$ leaving a tidal contribution of $\approx
-0.11\times 10^{-23}\rm rad\ s^{-2}$. This gives a rate of
$\frac{dR}{dt}\equiv \dot R=-0.15\times 10^{-9}\rm m\ s^{-1}$. Now
the apparent lunar and solar contributions amount to  $\approx
2.31\times 10^{-23}\rm rad\ s^{-2}$ and  $\approx 1.65\times
10^{-23}\rm rad\ s^{-2}$ respectively.  The most significant
result is that $\frac{dR}{dt}$ is negative and the magnitude is
about one tenth of the value derived using the tidal theory only.
Hence, Ghosh concluded that the Moon is actually approaching the
Earth with a vary small speed, and hence there is no
close-approach problem. Therefore, this will imply that the tidal
dissipation must have been much lower in the Earth's early
history.
\section{References}
Arbab, A. I., {\it Gen. Relativ. Gravit., \textbf{29}, 61, 1997}\\
Arbab, A. I., {\it Acta Geodaetica et Geophysica Hungarica,
\textbf{39}, 27, 2004}\\
Arbab, A. I., {\it Acta Geodaetica et Geophysica Hungarica,
\textbf{40}, 33, 2005}\\
 Gosh, A., {\it Origin of Inertia: extended Mach's principle and
cosmological consequences}, Apeiron, Canada (2000).\\
Stacey, F.,  {\it Physics of the Earth., Earth Tides}, J. Wiley \&
Sons Inc., New York, 1977.\\
Wells, J.W.,  \it{Nature}.\textbf{197}, 948 (1963)\\

\end{document}